\documentclass[letterpaper,twocolumn,useAMS,usenatbib,usegraphicx]{mn2e}
\usepackage[utf8]{inputenc}
\usepackage[UKenglish]{babel}
\usepackage{url}
\usepackage{amssymb}
\usepackage{aas_macros}

\newcommand{\hmpc}{$h^{-1}{\rm Mpc}$ }
\newcommand{\rnl}{\mathcal{R}(f_{\rm NL})}
\newcommand{\fnl}{f_{\rm NL}}

\newcommand{\lcdm}{$\Lambda$CDM }

\newcommand{\cardiff}{{School of Physics \& Astronomy, Cardiff University, 5 The Parade, Cardiff, CF24 3AA, United Kingdom}}

\pagerange{L20--L40}
\volume{418}
\pubyear{2011}
\date{Accepted 2011 August 16. Received 2011 August 16; in original form 2011 August }

\title[Exact EVT of HMFs]{Exact Extreme Value Statistics and the Halo Mass Function}
\author[Harrison \& Coles]{Ian Harrison\thanks{E-mail: ian.harrison@astro.cf.ac.uk} and Peter Coles\\ \cardiff \\}

\begin{document}
\maketitle
\begin{abstract}
Motivated by observations that suggest the presence of extremely
massive clusters at uncomfortably high redshifts for the standard
cosmological model to explain, we develop a theoretical framework for the
study of the most massive haloes, e.g. the most massive cluster
found in a given volume, based on Extreme Value Statistics (EVS). We
proceed from the \emph{exact} distribution of the extreme values
drawn from a known underlying distribution, rather than relying on
asymptotic theory (which is independent of the underlying form),
arguing that the former is much more likely to furnish robust
statistical results.  We illustrate this argument with a discussion
of the use of extreme value statistics as a probe of primordial
non-Gaussianity.
\end{abstract}
\section{Introduction}

The standard `concordance' or  Lambda cold dark matter ($\Lambda$CDM) cosmological model
incorporates the idea that large scale structure in the universe is
assembled hierarchically from Gaussian-distributed initial
perturbations in the density of Cold Dark Matter. In the
hierarchical models, structure in the universe forms in a `bottom
up' fashion, with small-scale density perturbations collapsing first
before merging over time to form larger and larger CDM haloes
\citep{White1991p, Peacock}. Baryonic matter falls into these
haloes, becoming shocked and virialised to form galaxies \citep[e.g.][]{Benson2010af}. The exact details of the rate and
magnitude of structure formation are highly sensitive to the
contents and dynamics of the universe and, as such, have the
potential to constrain  deviations from the minimal \lcdm model.
Indeed, the most massive collapsed object in the universe can
\emph{on its own} supply a definitive test of cosmological models, in
that the observation of a single sufficiently massive CDM halo has
the ability to rule out at high significance levels models in
which such a large object is unlikely to form. In particular, the
inference that extremely dense haloes must have arisen from large
upward density fluctuations seems a promising way to probe possible
departures from initial Gaussianity.

In accord with this line of reasoning, there has recently been considerable interest in the existence of high-mass, high-redshift
galaxy clusters as a means of identifying deviations from \lcdm
cosmology. Since the discovery by \cite{Jee2009a} of a
cluster at $z \sim 1.4$ with a mass of $8.5 \pm 1.7 \times 10^{14}
\mathrm{M_{\odot}}$, and other apparently challenging objects \citep{Brodwin2010, Santos2011, Foley2011}, several authors have reported tension between
the existence of such objects and concordance cosmology.
\cite{Jimenez2009}, \cite{Cay'on2011} and
\cite{Hoyle2011b} all report that this tension can be eased by the
presence of primordial non-Gaussianity, parameterised by $\fnl$, at
levels which far exceed (by a factor $\sim10$) the limits imposed by the CMB
\citep{Komatsu2011n}. Whilst models exist that predict a running
of $\fnl$ with scale \citep{LoVerde2008}, it is important to
explore the robustness of these detections before concluding that
changes to the standard model are needed. Furthermore, future
surveys will only increase  the observed volume in which clusters
may exist, so the most massive clusters found will increase
accordingly.

While the motivation for focussing on such objects is strong, in
order to perform model selection with high mass clusters we need to
understand the statistical properties of such objects. One way of
considering this problem is through Extreme Value Statistics (EVS)
\citep{Gumbel, KatzEVD}, which seek to make predictions for the
greatest (or least) valued random variable drawn from an underlying
distribution. There has recently been a resurgence of interest in
applying EVS to the field of cosmology with papers by
\cite{Mikelsons2009}, \cite{YamilaYaryura2010},
\cite{Colombi2011}, \cite{Davis2011},
\cite{Waizmann2011} and \cite{Chongchitnan2011}, the last
three dealing with high-mass clusters in particular. In this paper
we look more carefully at the underlying theory, derive from first
principles the exact extreme value statistics of the halo mass
function, and investigate their usefulness for constraining
cosmology.

The paper is organised as follows. In section \ref{sec:methods} we
introduce exact extreme value statistics and show how they may be
formulated for the case of the halo mass function, in both the \lcdm
case and one including amounts of primordial non-Gaussianity.
Section \ref{sec:results} compares this theoretical prediction for
the most-massive cluster with Monte-Carlo simulations. In section
\ref{sec:conclusions} we conclude and discuss prospects for
future work in this area.
\section{Methods}
\label{sec:methods}
\subsection{Exact and Asymptotic Extreme Value Statistics}

If we consider a sequence of $N$ random variates $\lbrace M_i
\rbrace$ drawn from a cumulative distribution $F(m)$ then there will
be a largest value of the sequence:
\begin{eqnarray}
\label{eqn:evs:max_seq} M_{\rm max} \equiv \sup \lbrace M_1, \ldots
M_N \rbrace.
\end{eqnarray}
If these variables are mutually independent and identically
distributed then the probability that all of the deviates are less
than or equal to some $m$ is given by:
\begin{eqnarray}
\label{eqn:evs:evs_cdf}
\Phi(M_{\rm max} \leq m;N) &=& F_1(M_1 \leq m)\ldots F_N(M_N \leq m) \nonumber \\
&=& F^{N}(m)
\end{eqnarray}
and the probability distribution for $M_{\rm max}$ is then found by
differentiating (\ref{eqn:evs:evs_cdf}):
\begin{eqnarray}
\label{eqn:evs:evs_exact}
\phi(M_{\rm max} = m;N) &=& N F^{\prime}(m) \left[ F(m) \right]^{N-1} \nonumber \\
&=& N f(m) \left[ F(m) \right]^{N-1}
\end{eqnarray}
This gives the \emph{exact} extreme value distribution for $N$
observations drawn from a known underlying distribution $f(m)$.
However, it is the seminal result of extreme value statistics
\citep{Frechet, FisherTippett} that, in analogy with the central
limit theorem for sample means, even in cases where $f(m)$ is not
explicitly known, in the limit $N \rightarrow \infty$ the
distribution $\phi(\hat{m}_N)$ of a suitably rescaled variable
\[
\hat{m}_N = \frac{m - a_N}{b_N},
\]
(where $a_N$ and $b_N$ are functions of $N$ determined by the
underlying distribution) asymptotically approaches one of only three
limiting forms: the Type-I, II and III (also known as Gumbel,
Fr\'{e}chet and Weibull respectively) extreme value distributions.
The functions $a_N$ and $b_N$ may be determined via the
\textit{reciprocal hazard function}:
\begin{eqnarray}
r(m) = \frac{1 - F(m)}{f(m)}
\end{eqnarray}
\begin{eqnarray}
b_N = F^{-1}\left(1 - \frac{1}{N}\right), \, a_N = r(b_N)
\end{eqnarray}

It is possible to encapsulate all these asymptotic distributions
within the Generalised Extreme Value (GEV) distribution:
\begin{eqnarray}
\label{eqn:gev}
G(\hat{m}_N; \gamma) &=& \exp\lbrace - \left[ 1 + \gamma \hat{m} \right] \rbrace^{-1/\gamma}_{+},
\end{eqnarray}
where values of the shape parameter $\gamma = 0$, $\gamma > 0$ and
$\gamma < 0$ pick out Type-I, II and III distributions respectively.
We have given this distribution the symbol $G(m)$ as opposed to
$\phi(m)$ to emphasize the difference between exact and asymptotic
distributions. It is possible to determine the asymptotic value of
$\gamma$ \citep{Gnedenko1943, Gyorgyi2010}, and hence
the asymptotic distribution type,  but this process proves to be
only analytically tractable for simple distributions.

The shape parameter describes the form of the asymptotic
distribution $G(\hat{m}_N; \gamma)$, but exact distributions
$\phi(m;N)$ will still have a best fitting value for $\gamma$.
Measuring $\gamma$ from a finite sized sample from a distribution
which is in the domain of attraction for the Type-I extreme value
distribution will lead to a measurement which converges towards zero
as the sample size increases. For distributions lying in the domain
of attraction of types II and III, $\gamma$ will converge to an
unknown value, depending on form of the underlying distribution. The
rate of this convergence can be spectacularly slow; for the specific
case of a Gaussian distribution (for which it can be analytically
determined that the asymptote is the $\gamma=0$ distribution)
convergence goes as $\sqrt{\ln N}$ only. It is therefore necessary
to be extremely careful that any observed value (or change in value)
of the shape parameter $\gamma$ is due to changes in the underlying
distribution, rather than due to the convergence of the exact
distribution $\phi(m;N)$ to the asymptotic one
$G(\hat{m}_N;\gamma)$.

\subsection{Extreme Value Statistics of the Halo Mass Function}

We now seek to determine the statistical distribution of extreme
values for the masses of CDM haloes, and in particular the validity
of the asymptotic form (\ref{eqn:gev}), for realistic cosmological
volumes. \cite{Press1974d} were the first to provide
an analytic method for predicting the co-moving number density
$n(M)$ of haloes of a given mass $M$, in differential form $dn/dM$,
considering spherical collapse of density perturbations in the
matter field. Subsequent to this, there has been much work
developing the halo mass function, both analytic and by fitting
functions to N-body simulations. We choose to use the mass function
from \cite{Sheth1999e} including effects from ellipsoidal
collapse:
\begin{eqnarray}
\label{eqn:hmfs:general_hmf}
    \frac{dn}{dM} =  A \sqrt{\frac{2a\delta_c}{\pi\sigma_M}} \exp{\left(-\frac{a \delta_c^2}{2\sigma_M^2}\right)} \left[ 1 + \left( \frac{\sigma_M^2}{a\delta_c^2}\right)^{p} \right] \frac{\bar{\rho}}{M} \frac{d \mathrm{ln}(\sigma_M^{-1})}{dM}.
\end{eqnarray}
Here, $\sigma_M^2$ is the variance of the matter field smoothed with
a top hat window of radius $R=(3M/4\pi\rho)^{1/3}$, with linear
power spectrum $P(k)$:
\begin{eqnarray}
\label{eqn:sigsquare}
    \sigma_M^2 = \int_{0}^{\infty} \frac{dk}{2\pi} \, k^2 P(k) W^2(k; R),
\end{eqnarray}
$\bar{\rho}$ is the mean density in the Universe, $\delta_c \simeq
1.686$ is the critical overdensity for collapse and $\lbrace A, a,
p\rbrace$ are parameters fitted to an N-body simulation and here
given their original values of $\lbrace0.322, 0.707, 0.3\rbrace$.
Throughout, we use a power spectrum calculated using
CAMB\footnote{\url{http://camb.info}} and the WMAP7+BAO+SN Maximum
Likelihood parameters from \cite{Komatsu2011n}. Using the halo
mass function as a predictor of number densities of haloes $n(M)$,
we can construct a probability distribution function (pdf) for halo
mass to be used in the calculation of the extreme value distribution
outlined above:
\begin{eqnarray}
\label{eqn:hmf_evs:evs_exact}
f(m) &=& \frac{1}{n_{\rm tot}}\frac{dn(m)}{dm}, \\
F(m) &=& \frac{1}{n_{\rm tot}}\left[ \int_{-\infty}^{M} dM \,
\frac{dn(M)}{dM} \right],
\end{eqnarray}
where the normalisation factor
\begin{eqnarray}
\label{eqn:evs_exact:norm} n_{\rm tot} = \int_{-\infty}^{\infty} dM
\, \frac{dn(M)}{dM}
\end{eqnarray}
is the total (co-moving) number density of haloes. For a constant
redshift box of volume $V$ the total number of expected haloes $N$
is then given by $n_{\rm tot}V$. These distributions can be inserted
into equation (\ref{eqn:evs:evs_exact}) to predict the pdf of the
highest mass dark matter halo within the volume.

\begin{figure}
 \begin{centering}
  \includegraphics[width=80mm]{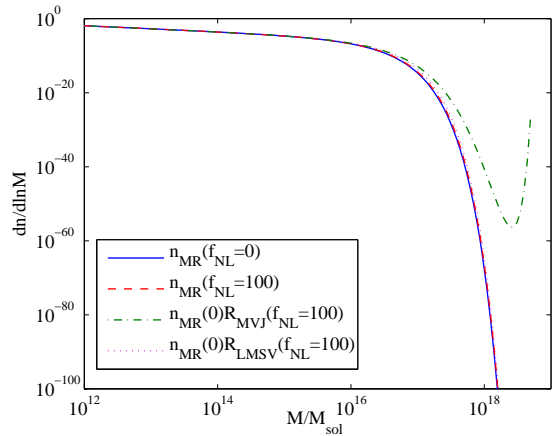}
 \caption{ Halo mass functions with non-Gaussianity applied using the prescriptions of \citet{Maggiore2010c} (MR), \citet{Matarrese2000} (MVJ) and \citet{LoVerde2008} (LMSV) showing the divergence of the MVJ prescription. }
    \label{fig:rnl_comp}
 \end{centering}
\end{figure}

The form of halo mass distribution in  \lcdm and alternative
cosmologies can also be examined; as an example of deviations from
\lcdm we include the effects of primordial non-Gaussianity. The halo
mass function has long been known to be sensitive to the presence of
primordial non-Gaussianity \citep{Lucchin1988a} and these effects
have been replicated within N-body simulations \citep{Grossi2009,
Pillepich2010a}. We include non-Gaussianity into the
model via the non-Gaussian correction factor $\rnl$ of
\cite{LoVerde2008} (LMSV):
\begin{eqnarray}
\label{eqn:hmf_evs:rlmsv}
\lefteqn{\mathcal{R}_{LMSV}(f_{NL}) =} \nonumber \\
& 1 + \frac{\sigma^2}{6\delta_c}\left[ S_3(\sigma) \left( \frac{\delta_c^4}{\sigma^4} - \frac{2\delta_c^2}{\sigma^2} - 1 \right) + \frac{d S_3}{d \ln\sigma}\left(\frac{\delta_c^2}{\sigma^2} - 1 \right) \right].
\end{eqnarray}
where $S_3$ is the normalised skewness of the matter density field, for which we use the approximation:
\begin{eqnarray}
\label{eqn:hmf_evs:s3}
S_3 \simeq 3 \times 10^{-4} f_{NL} \sigma^{-1}
\end{eqnarray}
given by equation (2.7) of \cite{Enqvist2011}. The choice of the
LMSV version is motivated by Figure \ref{fig:rnl_comp}, in which we
plot three methods of including primordial non-Gaussianity in the
halo mass function; the $\rnl$ correction factors of LMSV and
\cite{Matarrese2000} (MVJ) and the analytically applied
non-Gaussianity of \cite{Maggiore2010c} (MR), all applied to the
$\fnl = 0$ MR mass function. As can be seen (and as observed by \cite{Enqvist2011} when applied to the \cite{Tinker2008} mass function), the MVJ correction
factor leads to a divergence in the mass function in the high-mass
limit, which in this analysis we are still required to integrate
over. By applying non-Gaussianity to the MR mass function we can explicitly see that it is the $\rnl$ factor which leads to this divergence, rather than the mass function itself. In order to evaluate the efficacy of this formulation of the
extreme value statistics of the halo mass function,  we compare the
extreme value pdf calculated from
(\ref{eqn:hmf_evs:evs_exact}-\ref{eqn:evs_exact:norm}) to Monte
Carlo simulations of the most massive halo in a universe with a
given mass function. In each cosmology, we construct an ensemble of
realisations of the halo mass function; each realisation is
constructed by calculating the expected number of haloes in a bin of
width $\Delta \log m$ and drawing from a Poisson distribution with
this mean. The drawn value is then taken as the number of haloes in
this bin for this realisation, generating a mock catalogue of
uncorrelated haloes in the volume $V$. The largest cluster mass for
the realisation is determined as the central value of the highest
occupied bin (which is always singly occupied). The distribution of highest-mass cluster in each
catalogue is then recorded over $10^4$ realisations.
\section{Results and Comparisons with other work}
\label{sec:results}

\begin{figure}
 \begin{centering}
  \includegraphics[width=80mm]{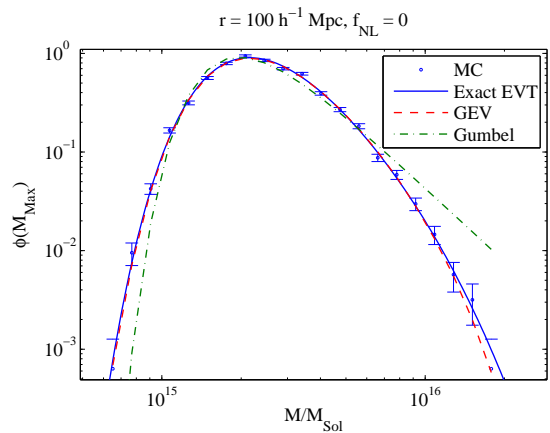}
  \caption{The extreme value distributions for the Sheth-Tormen halo mass function. Shown are the exact distribution and two best-fitting asymptotic distributions: a Type-I (Gumbel, dash-dotted) distribution and a general extreme value distribution with free $\gamma$ parameter (GEV, dashed).}
  \label{fig:hmf_dist}
 \end{centering}
\end{figure}

Figure \ref{fig:hmf_dist} shows the results of the above procedure
for the \cite{Sheth1999e} mass function with WMAP7 cosmological
parameters. Plotted are Monte Carlo results with Poisson errors, the
exact extreme value distribution calculated using
(\ref{eqn:evs:evs_exact}) and asymptotic Type-I (Gumbel) and GEV
distributions fitted using a maximum likelihood method. It can be
seen that the predictions of the exact extreme value distribution
(\ref{eqn:evs:evs_exact}) well match the results of the Monte-Carlo
simulations. As can be expected, including the extra degree of
freedom of the shape parameter $\gamma$ greatly improves the fit of
the GEV distribution over the Type-I.

\begin{figure}
 \begin{centering}
  \includegraphics[width=80mm]{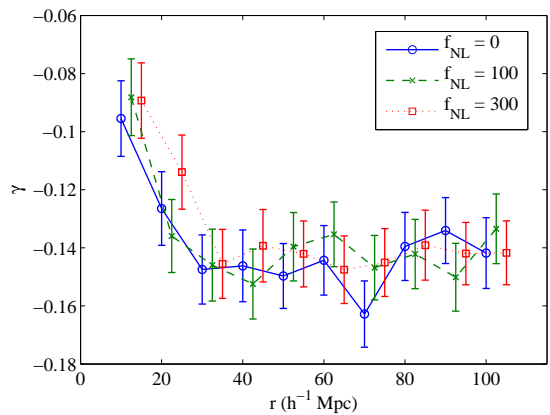}
  \caption{The shape parameter $\gamma$ for different volumes and values of $f_{NL}$, estimated using a maximum likelihood method and with $95\%$ error bars. Points for $\fnl=100$ and $\fnl=300$ are horizontally offset by $+2.5, +5$ \hmpc respectively. Convergence appears to be sufficient at volumes $\gtrsim30$ \hmpc and $\gamma$ appears to be poor at discriminating between different values of $f_{NL}$}
  \label{fig:convergence}
 \end{centering}
\end{figure}

Figure \ref{fig:convergence} shows the convergence of the shape
parameter $\gamma$ for a variety of spherical volumes and values of
the non-Gaussianity parameter $\fnl$. Values of $\gamma$ are
estimated with a maximum likelihood method and error bars represent
$95\%$ confidence intervals. As can be seen, whilst the shape
parameter appear well converged for volumes above $r\gtrsim30$
\hmpc, there is enough statistical noise so as to wash out any
potential detection of $\fnl \lesssim 300$ by using $\gamma$ as a
test statistic, even in this simple case with uncorrelated haloes.

\cite{Davis2011} also consider the extreme value statistics of
the halo mass function, forming the extreme value distribution as
the differential of the void probability:
\begin{eqnarray}
\label{eqn:hmf_evs:ddcsp_evt} \Phi^{\rm void}(M_{\rm max} = m) =
\frac{dP_{0}(m)}{dm}
\end{eqnarray}
where, in the Poisson limit, the void probability is given by:
\begin{eqnarray}
\label{eqn:hmf_evs:ddcsp_poisson}
P_0(m) = \exp ( -n(>m)V ).
\end{eqnarray}

\begin{figure}
 \begin{centering}
  \includegraphics[width=80mm]{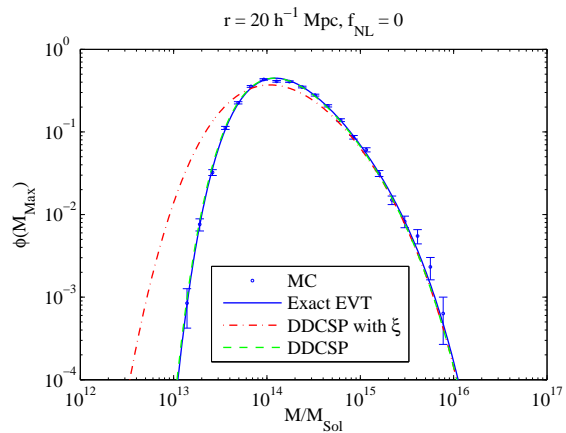}
 \caption{ Comparison of \citet{Davis2011} (DDCSP) and this work, showing the agreement of both methods of determining the extreme value statistics of the halo mass function. The dotted line represents the DDCSP version with halo correlations included. }
 \label{fig:ddcsp_comp}
 \end{centering}
\end{figure}

Shown in Figure \ref{fig:ddcsp_comp} is the comparison between the
extreme value distributions calculated using equations
(\ref{eqn:hmf_evs:ddcsp_evt}) and (\ref{eqn:evs:evs_exact}), showing
excellent agreement for the case of uncorrelated haloes, as is to be
expected. The method of \cite{Davis2011} can be readily modified
to account for correlated, biased haloes, primarily because of the
simple form taken by effects of correlations on the void
probability, but it remains a future endeavour to include these
effects in the exact model. However, the agreement of extreme value
distributions at the high mass end in the cases of both correlated
and uncorrelated haloes means that meaningful inferences on
likelihoods of most massive clusters may still be drawn from the
simple uncorrelated models.
\section{Discussion and Conclusions}
\label{sec:conclusions}

We have explored an avenue towards the construction of the
\emph{exact} distribution of halo masses which does not entail the
assumption that the distribution belongs to one of the asymptotic
types discussed in the classical literature of extreme value
statistics. Using both analytical and numerical techniques we have
shown that there can be significant differences between the exact
and asymptotic distributions and show in particular that the shape
parameter $\gamma$ is unlikely to provide an effective statistical
discriminator between Gaussian and non-Gaussian theories of
structure formation.

The approach we have taken relies on accurate knowledge of the
behaviour of the underlying distribution for large halo masses. Even
for the case of Gaussian initial conditions (i.e. $f_{\rm NL}=0$)
there is some theoretical uncertainty in what this behaviour
actually is. There exist a number of plausible halo mass functions in the literature \citep[e.g.][]{Sheth1999e, Jenkins2001n, Reed2003aa, Tinker2008}, all of which have differing tail behaviour and the level of indeterminacy worsens when we consider non-Gaussian models, as discussed in section \ref{sec:methods}.

Nevertheless, analytical approaches like those
discussed in this paper will certainly play an important role in
this area for some considerable time. The most massive haloes are so
rare that probing them using numerical techniques will require
enormous volumes to be simulated with sufficient resolution to
obtain accurate halo masses whilst at the same time avoiding
boundary artifacts. For example, in order to determine the
probability distribution of the most massive cluster in the Hubble
volume we would need an ensemble of simulations, each so large that
it would comprise a large number of independent Hubble volumes.
Faced with the significant computational cost of such a programme,
there can be no doubt that analytical theory, calibrated by smaller
scale simulations, will be the principal theoretical tool by which
extreme objects will be studied. We will adopt this approach in
future work.

The use of extreme value statistics as described in this work also has the advantage over studies which seek to use rare objects to constrain mass functions of clusters $n(M)$ \citep[e.g.][]{Vikhlinin2009, Allen2011} in that, in the EVS approach, a given object can always set a lower limit on the global extremum. This avoids the difficulty (in addition to the determination of cluster mass) of defining in a unbiased way precisely what volume is being probed, a process vulnerable to {\em a posteriori} selection effects.

\section*{Acknowledgments}
Ian Harrison receives funding from an STFC studentship. For the
purposes of this work Peter Coles is supported by STFC Rolling Grant
ST/H001530/1.


\end{document}